# Planar carbon nanotube-graphene hybrid films for high-performance broadband photodetectors

Yuanda Liu[1#], Fengqiu Wang[1#*], Xiaomu Wang[2*], Xizhang Wang[3], Emmanuel Flahaut[4,5], Xiaolong Liu[1], Yao Li[1], Xinran Wang[1], Yongbing Xu[1], Yi Shi[1], Rong Zhang[1*]

[1] School of Electronic Science and Engineering and Collaborative Innovation Center of Advanced Microstructures, Nanjing University, Nanjing 210093, China
[2] Department of Electrical Engineering, Yale University, New Haven, CT 06511, US
[3] School of Chemistry and Chemical Engineering, Nanjing University, Nanjing 210093, China
[4] CNRS; Institut Carnot Cirimat; F-31062 Toulouse, France
[5] Université de Toulouse; UPS, INP; Institut Carnot Cirimat; 118, route de Narbonne, F-31062 Toulouse cedex 9, France

**Abstract** Graphene has emerged as a promising material for photonic applications fuelled by its superior electronic and optical properties. However, the photoresponsivity is limited by the low absorption cross section and ultrafast recombination rates of photoexcited carriers. Here we demonstrate a photoconductive gain of ~$10^5$ electrons per photon in a carbon nanotube-graphene one dimensional-two dimensional hybrid due to efficient photocarriers generation and transport within the nanostructure. A broadband photodetector (covering 400 nm to 1550 nm) based on such hybrid films is fabricated with a high photoresponsivity of more than 100 $AW^{-1}$ and a fast response time of approximately 100 μs. The combination of ultra-broad bandwidth, high responsivities and fast operating speeds affords new opportunities for facile and scalable fabrication of all-carbon optoelectronic devices.

# Introduction

Combining low-dimensional nanomaterials into hybrid nanostructures is a promising avenue to obtain enhanced material properties and to achieve nanodevices operating with novel principles[1, 2]. The family of carbon allotropes, with its rich chemistry and physics, attracts a great deal of attentions in forming novel hybrid nanostructures[2]. In particular, the excellent electrical conductivities and large specific surface areas of 2D graphene and 1D carbon nanotubes (CNTs) have stimulated earlier theoretical and experimental investigations of 3D nanotube-graphene hybrid architectures for hydrogen storage[3], supercapacitors[4] and field-emitter devices[5]. Recently, ultra-thin CNT layers have been used as a reinforcing component for centimetre-sized chemical vapour deposition (CVD) grown graphene, where hybrid films with enhanced in-plane mechanical strength, uncompromised electrical conductivity and optical transparency are formed[6]. Even more remarkably, the synthesis of covalently-bonded single-wall carbon nanotubes (SWNTs) and graphene hybrid film is recently achieved and its use as flexible transparent electrodes is demonstrated[7]. The simple and scalable route for fabricating quasi-2-dimensional all-carbon hybrid films is envisioned to offer new opportunities beyond mechanical and nanoelectronic applications[2].

Both graphene and carbon nanotubes exhibit intriguing optical properties, such as broadband and tuneable light absorption, which make them promising materials for photodetectors[8, 9]. Graphene has proved excellent for ultrafast and ultrasensitive photodetectors. However, the relatively low absorbance of a single sheet of carbon atoms adversely limits the photoresponsivities of the earlier metal-graphene-metal devices (~$10^{-2}$ AW$^{-1}$)[10, 11]. A number of heterogeneous schemes employing plasmonic resonance[12, 13], microcavities[14], evanescent-wave coupling[15], and conventional semiconductor nanostructures[16] have been explored and lifted the photoresponsivity of graphene photodetectors to ~8.6 AW$^{-1}$ [13]. However, these approaches offer only modest responsivity enhancement and introduce fabrication steps that are not manufacturing scalable. Graphene photodetectors based on a photogating mechanism, i.e. by depositing a semiconducting quantum-dots (QDs) overlayer, is by far the most superior method in terms of responsivity enhancement and ease of fabrication[17, 18]. But the spectral coverage is limited by the absorption range of the QDs and enhancement is effective only at extremely low light intensities (i.e. pico-Watt or $10^{-12}$ W) due to the use of exfoliated graphene flakes (lateral size ~ 5 μm)[16]. Although an astonishing gain-bandwidth product of ~$10^9$ has been demonstrated[16], such systems are intrinsically slow with electrical bandwidth around 1-10 Hz, as limited by the low carrier mobility and long carrier recombination time of the QDs[17].

SWNTs, on the other hand, are π-conjugated, one-dimensional structures with nanometre diameters[19]. They exhibit either metallic or semiconductor features depending on the tube chiralities[20]. Both metallic and semiconducting SWNTs are effective light absorbers in a wide spectral range[21] and intrinsic mobility in semiconducting SWNTs is estimated to be as high as $10^5$ cm$^2$V$^{-1}$ s$^{-1}$ at room temperature[22]. Due to 1D quantum confinement, photoexcitations in SWNTs demonstrate rich physics[9]. For example, localised excitons with large binding energy is the main excitation for low-energy transitions (i. e. $S_{11}$) in semiconducting SWNTs[23], while metallic tubes can act as an efficient charge transport channel for SWNT ensembles to adjacent conducting media[24]. Although extensive efforts have been directed towards the realisation of nanotube photodetectors, the weak photoresponse (< $10^{-3}$ AW$^{-1}$) of single-tube photovoltaic devices greatly limits their practical use[25], while nanotube bolometric detectors are known to suffer from slow response time[26]. Recently, efficient charge transfer has been identified at junctions formed by graphene and SWNTs[24]. The intimate electronic coupling between the two sp$^2$-hybridized carbon allotropes, combined with the various strategies available for structural and chemical engineering of the interfacial electronic properties[27, 28], make such all-carbon hybrid an excellent candidate in enabling phototransistors with balanced and tuneable gain-bandwidth characteristics.

Here, we demonstrate a proof-of-concept photodetector based on a planar atomically-thin SWNT-graphene hybrid film. In our design, enhanced broadband light absorption is achieved in the hybrid film. Compared with pristine graphene photodetectors which exhibit only weak Schottky junction and high Auger recombination rate, the large built-in potential at the 1D-2D interface promotes effective separation of electron-hole pairs and reduces recombination of spatially isolated photocarriers. Furthermore, the trap-free interface enables a relatively fast operation rate. The devices exhibit a significant photoconductive gain of ~ $10^5$, together with a high electrical bandwidth of ~ $10^4$ Hz (response time ~ 100 μs) across visible to near-infrared range (400 nm – 1550 nm). The reported device constitutes a first implementation of large-area, quasi-2-dimensional SWNT-graphene hybrid film for optoelectronic devices and is envisaged to be important for optical communication, spectroscopy, remote sensing and high resolution imaging applications. Equally important, we demonstrate for the first time the possibility of harvesting robust excitons widely supported in 1D systems by a 2D layered material. Our results not only open up new avenues for studying fundamental carrier transport and relaxation pathways in nanometre-scale, 1D van der Waals junctions, but also pave the way for constructing high-performance optoelectronic heterostructures by planar 1D-2D hybrid building blocks, other than the use of purely 2D layered materials.

## Results

We fabricated proof-of-concept phototransistors using the SWNT-graphene hybrid film, as illustrated schematically in Figure 1a. The degenerately n-doped Si substrate with 285 nm thermal oxide was used as the back-gate. Wafer-scale SWNT-graphene hybrid film is fabricated by transferring CVD grown graphene onto an ultra-thin layer of SWNTs formed on the $SiO_2$/Si substrate (see Methods). Figure 1b,c shows a representative tapping mode AFM image of the hybrid film, where nanotubes form filament structures similar to the vein-like support as in ref 6. Figure 1d shows the height profile along the red line in Figure 1b, where unbundled SWNTs with diameters in the range of 1.0-1.6 nm are observed. This is in agreement with a mean tube diameter of ~1.4 nm as inferred from the RBM (Radial Breathing Mode) position of the Raman measurement (Supplementary Note 1 and Supplementary Figure 1). To investigate the physical characteristics of the 1D van der Waals junctions formed at graphene and SWNTs, a section of graphene is mechanically removed using a tape stripe (dark brown area, Figure 1c), so that the height of the SWNT-graphene junction with respect to the substrate can be directly measured. This yields a height of ~2.2 nm for the SWNT-graphene junction (Figure 1e), while the isolated SWNT height as measured from the uncovered portion indicates a height of 1.2 nm (Supplementary Figure 2). The height of graphene on $SiO_2$/Si substrate is measured to be ~0.8 nm, indicative of single atom layer[29]. The inferred offset at the SWNT-graphene interface is ~0.6 nm, which is larger than the interlayer distance of graphite (~0.335 nm), providing evidence for a long-range van der Waals interactions[30]. Figure 1f shows the optical absorption spectrum of the graphene and the SWNT-graphene hybrid film, illustrating enhanced broadband absorption due to the incorporation of SWNTs. The $S_{11}$ and $S_{22}$ bands are found to be located at ~1800 nm and ~1000 nm with higher SWNTs loadings (Supplementary Figure 3), which agree well with the tube diameter distributions.

To probe the electrostatic doping scenario at the 1D van der Waals junctions formed by graphene and SWNTs, transfer curves of the devices before and after forming the hybrid film are compared. It is observed that the Dirac point of the hybrid transistor shifted from 2 V to 17 V, indicating p-type doping of the graphene sheet by the SWNTs layer. It should be pointed out that due to the electronic inhomogeneity of the SWNTs, the electrostatic doping of graphene is expected be an overall effect from both metallic and semiconducting SWNTs. As metallic SWNT is found to form Ohmic contact with graphene[31] (i.e. there is no p-n junction or Schottky barrier), photoresponse of the hybrid channel is expected to arise from semiconducting SWNTs with suitable band alignment with graphene. Figure 1h provides a representative band alignment that supports the electrostatic doping scenario at the graphene semiconducting SWNT interface (where constituent semiconducting SWNTs have work functions larger than that of graphene), i.e. photogenerated electrons transport from SWNTs to graphene under the built-in field at the SWNT-graphene junction. In addition, the relatively low contact resistance for SWNT-graphene junctions (<0.1 MΩ)[31] as compared to those at SWNT-metal (1-4 MΩ)[32, 33] and SWNT-SWNT contacts (>2 MΩ)[34, 35] may further encourage photocarriers transport between SWNTs and graphene.

The real optical microscope image of the SWNT-graphene photodetector is shown in the inset of Figure 2a. No cracks or wrinkles are discernible. It is worth noting that in fabricating our device the SWNT layer is placed beneath graphene to ensure the formation of high quality SWNT-graphene interface, i.e. to avoid degradation caused by the residual PMMA used in the graphene transfer procedures[36]. Photoresponse of the SWNT-graphene hybrid device to 650 nm visible light is measured. Figure 2a shows a set of source-drain currents ($I_{SD}$) as a function of the back-gate voltage ($V_G$) under different illumination levels, where continuous negative shift of the voltage for Dirac point (the charge neutrality point) is observed, confirming a photocurrent generation mechanism as depicted by Figure 1h. Control devices with only SWNT layer as the channel are found to be insulating and exhibit no detectable photoresponse, probably due to the low density of nanotubes on the substrate (Inset of Figure 1g). The on/off ratio is about 4, suggesting graphene is the primary conductive channel[37]. For 650 nm light (1.9 eV) with photon energies larger than the $S_{22}$ bandgap of the semiconducting SWNTs (~ 1.2 eV), free electrons can be directly excited from the valence band to the conduction band. Photogenerated electrons in SWNTs are transferred to the graphene channel due to the built-in electric field, while the holes are trapped in the SWNTs due to the potential barrier at the interface. The conductance of the graphene channel decreases for $V_{BG}<V_D$ where the carrier transport is hole-dominated, but increases for $V_{BG}>V_D$, where carrier transport is electron-dominated. Figure 2b plots the negative shift of the Dirac point voltage ($\Delta V$) as a function of the illumination power, clearly revealing the high sensitivity of the SWNT-graphene device. $I_{SD}$ as a function of $V_G$ under different illumination power of 650 nm demonstrates effective tuning of the photocurrent and the photoresponsivity with the back-gate voltage, where a maximum photocurrent of > 90 µA is obtained at a negative $V_{BG}$ around -20 V (Supplementary Figure 4a). In figure 2c, linear scaling of the photocurrent with source-drain bias voltage ($V_{SD}$) for different optical powers at $V_{BG}=0$ is clearly observed. Figure 2b also plots the photoresponsivity as a function of the illumination power. To faithfully illustrate the performance of the device we define the responsivity using the incident power rather than the absorbed power. At an illumination power of ~0.2 µW, the photodetector exhibits a drastically enhanced responsivity of ~120 AW$^{-1}$, as compared with a bare graphene device ~ $10^{-2}$ AW$^{-1}$. Since the responsivity doesn't show sign of saturation, higher photoresponse on the order of ~1000 AW$^{-1}$ is expected at lower excitation power levels (Supplementary Figure 4b).

To rule out potential contributions from thermal effects, including bolometric contribution from the SWNT layer[26] and photothermoelelectric (PTE) contribution from the metal electrodes[38], we measured the $I_{SD}$-$V_G$ at temperatures from 53 K to 273 K (Supplementary Figure 5 and Supplementary Note 2). Source-drain current increases with an increasing temperature from 53 K to 173 K, and a slight decrease in the range 173 K-273 K, as expected for SWNT response[26]. The rather small temperature dependence of $I_{SD}$ allows us to rule out resistance increase caused by heating of SWNTs as the mechanism for photocurrent generation. Light induced heating of source or drain metal contacts is reported to lead to a temperature gradient, resulting in a photothermoelectric contribution to the photocurrent[38]. However, we did not observe detectable photocurrent in our devices by illuminating visible light (650 nm) on either metal contact. Therefore, the photoresponse of our device is ascribed to a photogating effect, where channel resistance is modulated through capacitive coupling caused by trapped photogenerated charges.

We further verify the operation mechanism of our devices by correlating the gate dependent photocurrents with the band alignment at the SWNT-graphene junctions. As schematically illustrated in Figure 2d, incident photons excite ground-state electrons of SWNTs into excited states, and then electron-hole pairs are formed at the SWNT-graphene interface. For $V_G$<0, the Fermi level of graphene is lowered, which leads to a steeper upward band bending and an enhanced built-in electric field at the SWNT-graphene junction. This facilitates more photoelectrons to transfer from SWNTs to graphene, leading to an increasing photocurrent. As the magnitude of negative $V_G$ continues to increase, the trapped holes in the valence band of SWNTs begin to tunnel through the thinned barrier into the graphene channel, resulting in a decreased photocurrent. In the case of $V_G$ >0, the Fermi level of graphene is raised up, the potential offset for electrons is effectively decreased. Therefore, the photocurrent at a given positive gate voltage is smaller than that for a negative gate voltage of equal amplitude. When $V_G$ is sufficiently high, the SWNT energy band will flip to bend downwards, allowing photogenerated holes to transfer to the graphene channel. This is the reason why the conductivity decreases for $V_G$>40 V, and the photocurrent subsequently switches sign (Supplementary Figure 4a).

Response time is another key figure of merit for photodetectors and is also relevant in revealing the physical mechanism of the device operation. Figure 3a shows the on/off source-drain current of the photodetector at an incident power of 440 µW (650 nm, $V_{SD}$=1.2 V). The temporal photoresponse was measured at 173 K to suppress the outside interference including scattering centres from the substrate and charge-trapping surface states[31]. After hundreds of on-off cycles, photocurrent level is well retained, demonstrating good reliability and reversibility of our devices. At the same experimental conditions, we did not observe any detectable photocurrent in a control device with a bare graphene channel. Figure 3b gives the high-resolution temporal response (sampling interval 100 µs). In order to encourage photocarriers separation we apply a 5 V source-drain voltage. Sharp responses are observed from which we estimate the rise time and the fall time to be ~ 100 µs. The operation speed of our device is more than 1000 time faster than a graphene-QDs detector with similar photosensitive area[17], and can be further reduced by scaling the device dimensions. The high mobility and fast transfer of carriers within the SWNT-graphene hybrid film are believed to account for the excellent temporal performance of our devices.

The EQE (External Quantum Efficiency) was estimated using the parallel plate capacitor model[18], QE $=\Delta V \times C_{ox}/\phi_{photon}$, where $\Delta V$ is the Dirac voltage shift, $C_{ox}$ is the capacitive coupling of the backgate for 285 nm-thick thermal oxide ($7 \times 10^{10}$ cm$^{-2}$ ·V$^{-1}$), $\phi_{photon}$ is the photon flux. As shown in Figure 3c, for 650 nm light our devices show a power-dependent EQE ranging from 34% to 12% for power illumination <1 μW, and an EQE of <0.7% with incident power >25 μW. The peak value for the QE is slightly higher than the 25% achieved with QDs[18], demonstrating the high efficiency of photocarriers generation and transport in the SWNTs-graphene hybrid. From the transfer curve measured in the absence of light, it is estimated by the Kim model[39] that the field-effect mobility for electrons and holes are 3920 and 3663 cm$^2$V$^{-1}$ s$^{-1}$, respectively. This is superior to graphene sensitized with PbS quantum dots[18]. Such high mobilities confirm that graphene can be doped by 1D SWNTs, yet preserve its outstanding electronic properties, facilitating high gain and fast response time. The carrier transit time of our device is estimated to be on the order of 10$^{-9}$ s (based on a carrier mobility of ~3920 cm$^2$V$^{-1}$ s$^{-1}$, a channel length of 70 μm and a bias of 5 V), thus the photoconductive gain $G=\tau_{lifetime}/\tau_{transit}$ (the ratio of the lifetime of the trapped holes over the electron transit time from source to drain) is on the order of ~10$^5$ using a trapped carrier lifetime of ~100 μs. Therefore the gain-bandwidth product of our SWNT-graphene device is on the order of $1 \times 10^9$ Hz (given by the product of ~10$^5$ gain and ~10$^4$ Hz of bandwidth), which is quantitatively similar to the PbS QD-graphene hybrid photodetector[18].

Characteristics of broadband photoresponse of our device over a range of incident wavelengths is investigated by using multiple laser diodes operating from visible to the near infrared range (405, 532, 650, 980 and 1550 nm). The responsivities for various wavelengths are summarized in Figure 3d, where ultra-broadband spectral coverage is observed. Higher responsivities were obtained at lower incident power levels. At a moderate illumination power of ~0.3 μW, all wavelengths exhibit a responsivity > 10 AW$^{-1}$, a typical specification for commercial photodetectors. The responsivity value is higher for shorter wavelengths than for longer ones (Supplementary Figure 6), consistent with the absorption spectrum of the SWNTs used, which is the usual case for phototransistors based on a photogating mechanism[18]. From the wavelength dependence, even higher photoresponsivities are expected for ultraviolet wavelengths.

**Discussion**

The essential ingredients of our devices are the sensitivity to external electrostatic in graphene conductivity, the broadband and tuneable photon absorption in both graphene and SWNTs, and the high speed transportation of photocarriers in the planar SWNT-graphene hybrid. To resolve the impact tube diameters and chiralities may have on the photodetection processes, we fabricated control devices using highly purified metallic tubes and (6, 5) chirality enriched semiconducting SWNTs. Quite different photodetection mechanisms were revealed (Supplementary Figure 7 and 8, Supplementary Note 3). Figure 3e compares the photoresponsivities of purified metallic and semiconducting SWNTs based devices, and the responsivity of the metallic tube based devices is seen to be ~ 5% of that of (6, 5) chirality tube based devices. The results clearly identify semiconducting tubes as the main contribution for photocurrents. In the meantime, unlike nanoelectronic devices where distinct SWNT chirality is desirable, our devices will not degrade by the presence of multiple species of SWNTs, similar to SWNT based nonlinear absorption devices[40]. The mitigation of requirements for tube chiralities also facilitates the practical and scalable fabrication of such films. The performance of the proof-of-concept photodetector based on SWNT-graphene hybrid film can be further optimized by multiple strategies including controlling the density, chiralities[41] and alignment direction of the nanotube layer and by using refined SWNT-graphene hybrid film fabrication methods[6,7].

In summary, by combining large-area CVD grown graphene with atomically-thin SWNT layer, we have formed a quasi-2-dimensional all-carbon hybrid film, which exhibits strikingly enhanced photodetection capabilities superior to either graphene or SWNTs. In contrast to previous charge trap based phototransistors, such hybrid film based photodetectors exhibit not only a significant photoconductive gain of ~$10^5$, but also a fast response time of ~100 μs and an ultra-broadband sensitivity across visible to near-infrared, covering the telecommunication band at ~1.5 μm. Benefiting from the solution processability of SWNTs, large-scale growth and transfer of graphene, as well as compatibility with standard photolithography, our devices exhibit great potential for use in photodetection applications requiring large sensing area, high photoresponsivities, video-frame-rate processing and substrate flexibility. The results demonstrated here represent a significant step towards facile and scalable fabrication of high-performance optoelectronic devices using all-carbon hybrid nanostructures and have substantial implications for fundamental investigation of the van der Waals interactions between layered materials and in-plane quantum wires, in the one-dimensional limit.

## Methods

**Phototransistor fabrication.** The graphene samples were grown on copper foil by CVD method, and Raman spectroscopy combined with optical microscope characterizations point to a defect-free single-layer sample. We use single-wall carbon nanotubes (SWNTs) from a commercial supplier (Carbon solutions Inc.). The phototransistors are fabricated as follows: SWNTs suspensions are produced by tip-sonicating 1 mg nanotube soot in 10 mL NMP (N-methyl-2-pyrrolidone). The resultant suspensions are centrifuged with 10,000 g for 1h before the supernatant is collected for the fabrication of SWNTs thin film on $SiO_2$/Si wafer. CVD graphene is transferred on top of the SWNTs layer using the PMMA supported procedures. Electrodes are patterned by standard photolithography. Different metal composition (Ti/Au and Pd/Au) are subsequently deposited. The graphene channel is patterned by another photolithography and oxygen plasma etching.

**Photoresponse characterisation.** For photoresponse characterization, we used 405, 532, 650, 980, and 1550 nm laser diodes, respectively. The beam is guided through an optical fiber with a FC/PC ferrule and is subsequently incident onto the channel of the devices without focusing. The beam at the device was measured to be Gaussian-shaped with a diameter of about 300 μm at 650 nm illumination. The area of the channel is less than 100 μm×40 μm. Therefore, when calculating the photoresponsivity, we used only the portion of the incident light intensity that overlaps with the channel area. The electrical measurements were carried out in a closed cycle cryogenic probe station under vacuum ($10^{-6}$ Torr) at room-temperature and the data were collected by a Keithley-4200 semiconductor parameter analyzer. The temporal photoresponse was measured at 173 K for suppressing the outside interference.

**AFM measurements.** AFM measurements were performed using an Asylum Research Cypher AFM operating at room temperature and ambient conditions.

**Raman and optical absorption spectra.** Raman measurements were performed on a Horiba Jobin Yvon LabRAM HR 800 system using a 514 nm excitation laser operating at 1 mW, 100X objective lens with about 1 μm diameter spot size, and 1800 lines/mm grating with about 0.45 $cm^{-1}$ spectral resolution. Optical absorption spectrum was measured using a Shimadzu UV-vis-NIR UV-3600 spectrophotometer.

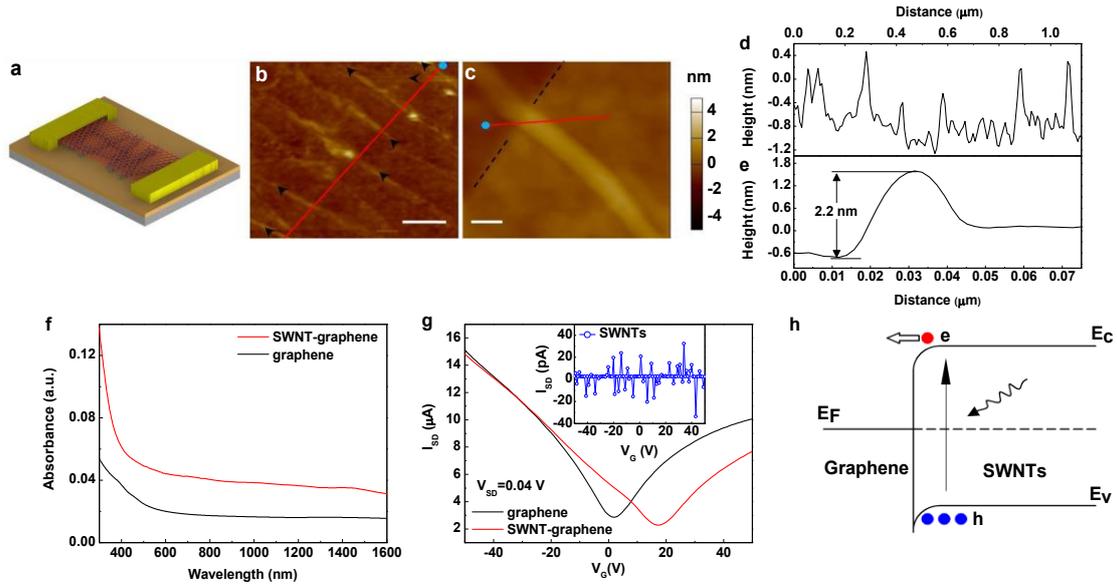

**Figure 1. Planar SWNT-graphene hybrid film and photodetector. a**, Schematic of the phototransistor. **b**, AFM image of the hybrid film on the SiO$_2$/Si substrate (scale bar, 200 nm). **c**, AFM image of one individual SWNT partially covered by graphene (scale bar, 20 nm). SWNT was uncovered by mechanically exfoliating the top graphene using a tape stripe. The black dashes show the edge of the graphene (left: SiO$_2$; right: graphene). **d**, Height profile along the red line in panel b. The blue dot in panel b marks the zero point in panel d. **e**, Comparison of the respective height profiles of the graphene-covered portion and the uncovered portion of SWNT shown in panel c. **f**, UV-vis-IR absorbance curves of graphene and SWNT-graphene hybrid film on quartz. **g**, Transfer characteristics of the graphene, SWNTs and SWNT-graphene transistors without light. Comparing with the graphene transistor, the Dirac point of the SWNT-graphene transistor shifted from 2 V to 17 V, indicating p-type doping in the graphene sheet induced by SWNTs. The electron (hole) mobility decreases from 6142 cm$^2$V$^{-1}$ s$^{-1}$ (7146 cm$^2$V$^{-1}$ s$^{-1}$) to 3771 cm$^2$V$^{-1}$ s$^{-1}$ (4666 cm$^2$V$^{-1}$ s$^{-1}$). The inset exhibits the source-drain current of the SWNTs transistor is less than 10$^{-11}$ A, indicating disconnected electric pathways for the pristine SWNTs channel. **h**, The energy band diagram at the junction formed by graphene and semiconducting SWNTs. Photogenerated electrons in SWNTs are transferred to graphene due to the built-in field at the junction.

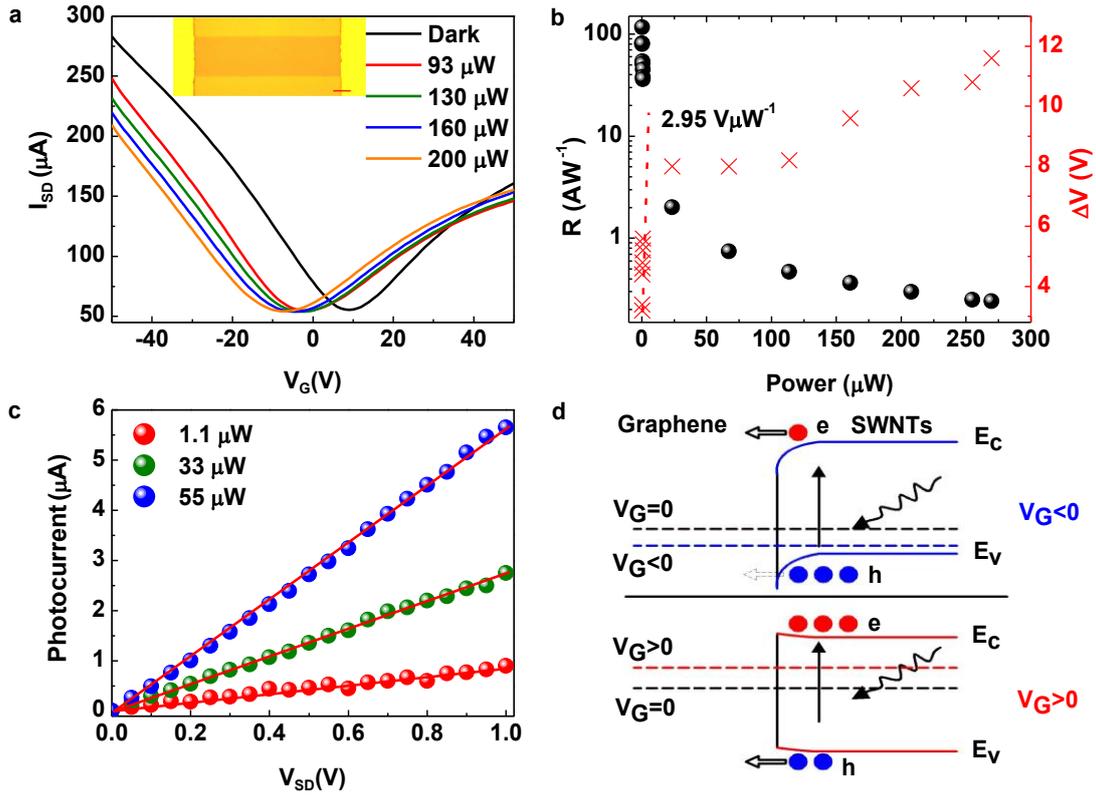

**Figure 2. Photoresponse performance under 650 nm illumination. a**, Source-drain current ($I_{SD}$) as a function of back-gate voltage ($V_G$) for the SWNT-graphene device with increasing 650 nm illumination intensities. $V_{SD}$=0.5 V. Increasing of the illumination leads to a photogating effect that shift the Dirac point to lower $V_G$, indicating electron doping of the graphene sheet. The inset is the optical micrograph of the fabricated device (scale bar, 10 μm). The gold areas indicate the metal electrodes. The orange area and LT orange areas are graphene channel and $SiO_2$/Si substrate, respectively. **b**, Responsivity (black dots) and shift of Dirac point (red cross) as a function of the 650 nm illumination. **c**, The magnitude of the photocurrent increases linearly with source-drain bias voltage ($V_{SD}$) for different optical powers ($V_{BG}$=0). Red lines are linear fits. **d**, Energy-band diagram of the SWNT-graphene phototransistors at different back-gate voltage. The dash lines correspond to the Fermi level of graphene at different gate voltages. The blue and red lines schematically illustrate the gate voltage dependence of the SWNT energy levels.

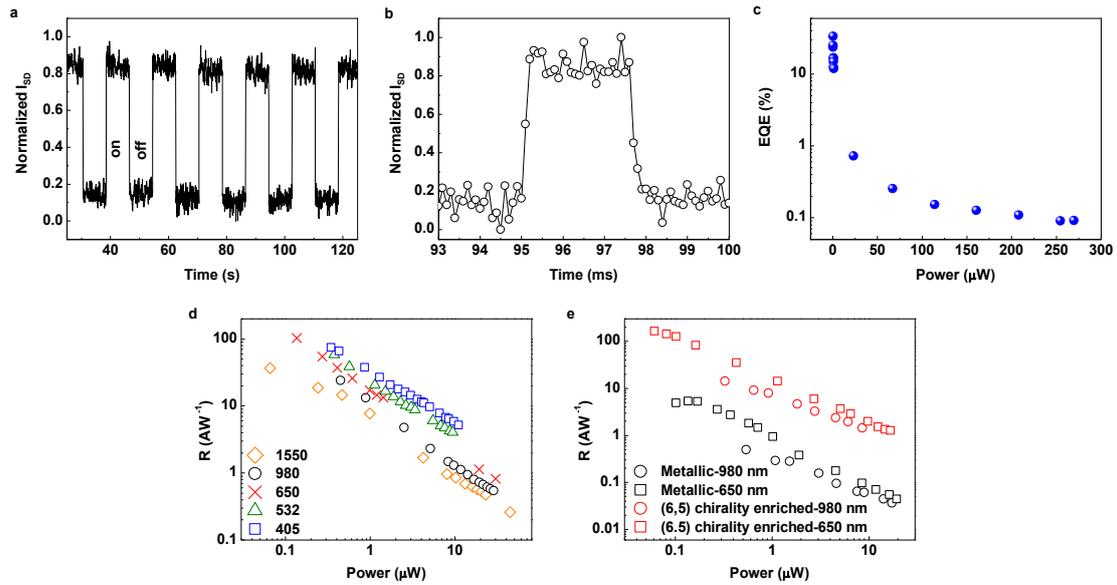

**Figure 3. Temporal characteristics and ultra-broadband photoresponse. a**, **b**, Temporal photocurrent response of the SWNT-graphene hybrid photodetector, indicating a rise time and a fall time on the order of ~100 μs. The illumination power is 440 μW and the laser wavelength is 650 nm. **c**, External quantum efficiency as a function of illumination power at 650 nm. **d,** Responsivities as a function of the optical power for different illumination wavelengths (405, 532, 650, 980 and 1550 nm). **e,** Comparison of responsivities measured from devices using purified metallic and (6,5) chirality enriched SWNTs. It is shown that the responsivity of the metallic tube based devices is ~ 5% of that of (6, 5) chirality tube based devices.

## Acknowledgements

This work was supported in part by National Key Basic Research Program of China 2014CB921101, 2011CB301900, 2013CBA01604; National Natural Science Foundation of China 61378025, 61450110087, 61427812, 61274102, 61325020, 61261160499, 61504056; Natural Science Foundation of Jiangsu Province BK20140054. Jiangsu Province Shuangchuang Team Program. Y.D.L. acknowledges funding of the China Postdoctoral Science Foundation 2014M551558 and Jiangsu Planned Projects for Postdoctoral Research Funds 1402028B.

## Author contributions

F.W., X.M.W. and R.Z. conceived and supervised the project. Y.D.L., Y.L. and X.Z.W. performed the photocurrent measurements. Y.D.L. and E.F. carried out sample preparation and device fabrication. X.L. and X.R.W. performed the AFM measurement. F.W., X.M.W., Y.X, Y.S., and R.Z. co-wrote the paper with all authors contributing to discussion and preparation of the manuscript.

[#] These authors contributed equally to this work.
Emails: fwang@nju.edu.cn; xiaomu.wang@yale.edu; rzhang@nju.edu.cn